\documentstyle[preprint,aps,epsf]{revtex}

\def\singlespace {\smallskipamount=3.75pt plus1pt minus1pt
                  \medskipamount=7.5pt plus2pt minus2pt
                  \bigskipamount=15pt plus4pt minus4pt
                  \normalbaselineskip=12pt plus0pt minus0pt
                  \normallineskip=1pt
                  \normallineskiplimit=0pt
                  \jot=3.75pt
                  {\def\smallskip {\vskip\smallskipamount}}
                  {\def\medskip   {\vskip\medskipamount}}
                  {\def\bigskip   {\vskip\bigskipamount}}
                  {\setbox\strutbox=\hbox{\vrule
                    height10.5pt depth4.5pt width 0pt}}
                  \parskip 7.5pt
                  \normalbaselines}
\def\middlespace {\smallskipamount=5.625pt plus1.5pt minus1.5pt
                  \medskipamount=11.25pt plus3pt minus3pt
                  \bigskipamount=22.5pt plus6pt minus6pt
                  \normalbaselineskip=22.5pt plus0pt minus0pt
                  \normallineskip=1pt
                  \normallineskiplimit=0pt
                  \jot=5.625pt
                  {\def\smallskip {\vskip\smallskipamount}}
                  {\def\medskip   {\vskip\medskipamount}}
                  {\def\bigskip   {\vskip\bigskipamount}}
                  {\setbox\strutbox=\hbox{\vrule
                    height15.75pt depth6.75pt width 0pt}}
                  \parskip 11.25pt
                  \normalbaselines}
\def\doublespace {\smallskipamount=7.5pt plus2pt minus2pt
                  \medskipamount=15pt plus4pt minus4pt
                  \bigskipamount=30pt plus8pt minus8pt
                  \normalbaselineskip=30pt plus0pt minus0pt
                  \normallineskip=2pt
                  \normallineskiplimit=0pt
                  \jot=7.5pt
                  {\def\smallskip {\vskip\smallskipamount}}
                  {\def\medskip   {\vskip\medskipamount}}
                  {\def\bigskip   {\vskip\bigskipamount}}
                  {\setbox\strutbox=\hbox{\vrule
                    height21.0pt depth9.0pt width 0pt}}
                  \parskip 15.0pt
                  \normalbaselines}

\begin{document}
\preprint{
\hfill$\vcenter{\hbox{\bf IUHET-405} \hbox{March
             1999}}$  }

\title{\vspace*{.75in}
Leptogenesis and Yukawa textures}

\author{Micheal S. Berger
\footnote{Electronic address:
berger@gluon.physics.indiana.edu} and Biswajoy Brahmachari 
\footnote{Electronic address: biswa@gluon2.physics.indiana.edu}}

\address{
Physics Department, Indiana University, Bloomington, IN 47405, USA}

\maketitle

\thispagestyle{empty}

\begin{abstract}
We study a set of textures giving  rise to correct masses
and mixings of the charged fermions in the context of leptogenesis. The
Dirac neutrino texture pattern is assumed to be identical with the up quark
texture. The heavy Majonana neutrino mass matrix is obtained by inverting
the type-I see-saw formula and using the neutrino  masses and mixings
required by the solar and atmospheric neutrino oscillation experiments as
input. After making the feasibility study of the generated lepton
asymmetry  via the decay of the heavy right handed neutrino, we compute
the generated baryon asymmetry by numerically solving the supersymmetric
Boltzmann equations. We find for these models that both the hierarchy of 
the texture as well as the placement of the texture zeros are important 
to the viability of leptogenesis as the source of the observed baryon 
asymmetry of the universe.

\end{abstract}

\newcommand{\be}{\begin{equation}}
\newcommand{\ee}{\end{equation}}
\newcommand{\bea}{\begin{eqnarray}}
\newcommand{\eea}{\end{eqnarray}}

\newpage

\section{Introduction}

Relationships between fermion masses and mixings have been the subjects of 
much theoretical interest starting with a postulated relationship between
the Cabibbo angle and the down and strange quark masses. Most of the
unknown parameters in the Standard Model (SM) occur in the 
Yukawa sector and any relationships between these parameters is welcome
theoretically and experimentally testable. The interest in models
of fermion masses and mixings accelerated with the advent of grand unified
theories. In these models, the gauge multiplets of the Standard Model are 
unified into multiplets of the grand unified gauge group, and relationships
between the parameters emerge naturally as a consequence of the larger 
symmetry. These models can be augmented by global symmetries or texture zeros
in the Yukawa coupling matrices can be assumed, leading to further predictions.

A mystery of the Yukawa sector is the obvious hierarchy that exists
in fermion masses and mixings. The 
top quark mass is much larger than the charm quark mass which is still much
larger than the up quark mass, for example. The (Cabbibo-Kobayashi-Maskawa)
CKM matrix is measured experimentally have
small mixing angles. Clearly a fundamental theory that explains the origin
of the couplings in the Yukawa sector rather than just parameterizing
them, should explain these features. The Yukawa sector of the Standard
Model is parametrized in terms of $3\times 3$ matrices, so the hierarchy
exhibits itself as a hierarchy among the elements of these matrices.

The usual predictions from these models of fermion masses and mixings are
relationships between the masses of quarks and leptons or between the mixing
angles of the CKM matrix and dimensionless ratios of the quark and lepton 
masses. In addition there are often predictions for the amount of 
charge conjugation-parity (CP) violation
and predictions for the CP asymmetries of meson decays. It is of interest
to consider whether further constraints are obtained after making reasonable
assumptions about some other physical observable. In this paper we
consider textures (or patterns of zero entries) of Yukawa
coupling matrices and assume that the baryon asymmetry has its
origins in the decay of heavy Majorana neutrinos which violate lepton
number. The asymmetry in lepton number is recycled into a baryon number
asymmetry via the sphaleron process. This idea was first put forward by
Fukugita and Yanagida\cite{fy}
and has come to be known as baryogenesis via leptogenesis. While this 
constraint
is admittedly more speculative than the comparison of masses and mixing angles
derived from experiment, it is instructive to determine which properties of 
the Yukawa textures are essential for the baryogenesis via leptogenesis to 
work.

Sakharov pointed out\cite{Sakharov} 
that a small baryon asymmetry may have been produced in 
the early universe if three conditions are satisfied: 1) baryon number is 
violated, 2) charge conjugation symmetry (C) and CP are violated, and 3) there
is a departure from thermal equilibrium\footnote{A baryon asymmetry could 
arise even in thermal equilibrium if CPT is violated. See, for example, 
Ref.~\cite{bckp}.}. Since both C and CP are violated in
the Standard Model, and baryon number is violated by a nonperturbative effect
called sphalerons, the natural place to look first to explain the generation
of a baryon asymmetry is within the Standard Model itself. However this 
line reasoning does not work: the required Higgs masses is too small and has
been ruled out by the direct searches at LEP\cite{SM-bary}. 
One can try to extend the Standard Model: one popular attack is to consider
the Minimal Supersymmetric Standard Model (MSSM) and assume that the source
of the CP violation is still contained within the CKM matrix\cite{MSSM-bary}.
One can achieve the observed baryon asymmetry, but only at the expense of 
going to a corner of parameter space, namely one requires a light scalar top 
quark (top squark or stop). This kind of solution should rightly be regarded
as unnatural on theoretical grounds, although it is of much experimental 
and phenomenological interest primarily because it is just out of reach.

We pursue in this paper the alternative approach
of baryogenesis via leptogenesis: With the confirmation of the 
atmospheric neutrino anomaly in the SuperKamiokonde\cite{super}, it seems 
plausible to start from the point of view that it is likely that the 
neutrino mixing is occuring and the neutrinos have masses given by a 
see-saw mechanism. 
The heavy Majorana neutrino mass matrix is obtained by inverting the
type-I see-saw formula 
\be
m_{\rm eff}=m^T_D { 1 \over M_N} m_D \label{eqn1}\;,
\ee
and using the neutrino masses and mixings required
by the solar and atmospheric neutrino oscillation experiments as 
input\footnote{We neglect the possibility that there are contributions 
from a left-handed triplet Higgs}.
The neutrino sector contains a new source of CP violation; the interference
between tree-level and one-loop contributions to the Majorana neutrino decays
can give rise to a lepton asymmetry.
In this scenario, the amount of CP violation that gives rise to a lepton
asymmetry and ultimately to a baryon asymmetry depends critically on the 
Dirac mass matrix of the neutrinos in two ways: (1) the overall hierarchy 
pattern of the matrix and (2) the placement of the texture zeros\footnote{If 
a texture zero predicts a small level of CP violation at the GUT scale, this 
suppression will be preserved by the renormalization group scaling.}. 
Furthermore, the generated lepton asymmetry can be erased by subsequent
lepton-number violating scattering and this dilution can depend on the 
placement of the texture zeros. 

In this paper we start from the position that the positive observation in 
the solar and atmospheric neutrino experiments suggest that there is a 
new scale of heavy physics. We assume that heavy right-handed neutrinos exist
and the lightness of the observed neutrinos is the result of a seesaw 
mechanism. In this framework we study the plausibility of leptogenesis 
in the case of neutrino Dirac mass matrices 
with texture zeros and hierarchical structure similar to the ones 
that are consistent with low-energy data in the quark sector. 

\section{Yukawa Textures}

\begin{table}[htb]
\begin{center}
\[
\begin{array}{|c||c||c|}
\hline
Solution &  \lambda_u & \lambda_d \\ 
\hline
&&\\
1 
& 
\pmatrix{0 & \sqrt{2} \lambda^6 & 0 \cr
            \sqrt{2} \lambda^6 & \lambda^4 & 0 \cr
            0 & 0 & 1 }
& 
\pmatrix{0 & 2 \lambda^4 & 0 \cr
          2 \lambda^4 & 2 \lambda^3 & 4 \lambda^3 \cr
            0 & 4 \lambda^3  & 1 }

\\
&&\\
2
&
\pmatrix{0 & \lambda^6 & 0 \cr
            \lambda^6 & 0 & \lambda^2 \cr
            0 & \lambda^2 & 1 }
&
\pmatrix{0 & 2 \lambda^4 & 0 \cr
          2 \lambda^4 & 2 \lambda^3 & 2 \lambda^3 \cr
            0 & 2 \lambda^3  & 1 } 

\\
&&\\
3
&
\pmatrix{0 & 0 & \sqrt{2} \lambda^4 \cr
         0 & \lambda^4 & 0 \cr
            \sqrt{2} \lambda^4 & 0 & 1 }
&
\pmatrix{0 & 2 \lambda^4 & 0 \cr
          2 \lambda^4 & 2 \lambda^3 & 4 \lambda^3 \cr
            0 & 4 \lambda^3  & 1 } 

\\
&&\\
4
&
\pmatrix{0 & \sqrt{2} \lambda^6 & 0 \cr
            \sqrt{2} \lambda^6 & \sqrt{3} \lambda^4 & \lambda^2 \cr
            0 & \lambda^2 & 1 }
&
\pmatrix{0 & 2 \lambda^4 & 0 \cr
          2 \lambda^4 & 2 \lambda^3 & 0 \cr
            0 & 0  & 1 } 

\\
&&\\
5
&
\pmatrix{0 & 0 & \lambda^4 \cr
         0 & \sqrt{2} \lambda^4 & \lambda^2/\sqrt{2} \cr
            \lambda^4 & \lambda^2/\sqrt{2}  & 1 } 
&
\pmatrix{0 & 2\lambda^4 & 0 \cr
         2\lambda^4 & 2\lambda^3 & 0 \cr
           0 & 0  & 1 } 

\\
&&\\
\hline
\end{array}
\]
\end{center}
\caption{The list of textures suggested by RRR.} 
\label{table1}
\end{table}

Ramond, Ross and Roberts (RRR) performed\cite{rrr} 
a systematic search for all possible symmetric quark and lepton mass
matrices with five texture zeros at the unification scale that are compatible
with low-energy measurements. They found a total of
five possible solutions, which we display again in Table~\ref{table1} for
convenience.

We assume the Dirac neutrino mass matrix at the GUT scale has the same 
texture zeros as the up quark matrix
\begin{equation}
m_D^\prime \simeq m_u =  \lambda_u ~v~\sin \beta.
\end{equation}
In certain
situations where the Yukawa interactions are minimal,
grand unified symmetry enforces an exact equality. More generally one might 
expect the equality of elements in the neutrino texture and the up quark 
texture not be exact, but be related by Clebsch coefficients (very often 
3). These Clebsch factors are typically small and do not upset the hierarchy
of the matrices. 
The general qualitative features of our analysis is not affected by 
these factors of order one, since the amount of baryon asymmetry generated in
a model with a particular texture is governed by the hierarchy (given in terms
of the parameter $\lambda$ which is fixed by the Cabibbo angle)
and the position of the texture zeros. 

\section{Numerical Solutions}
We consider small angle MSW solution of the solar neutrino problem
through the mixing $\nu_e \leftrightarrow \nu_\mu$ and maximal mixing
solution of atmospheric neutrino oscillation through the mixing
$\nu_\mu \leftrightarrow \nu_\tau$. We take as inputs the
following neutrino masses consistent with the experimental  
measurements, namely\cite{super,constraints}
\bea
0.8 < \sin^2 2 \theta_{23} < 1\;, &&\quad 
10^{-3} < \Delta m_{23}^2 <10^{-2}\;, \\
3 \times 10^{-3}< \sin^2 2 \theta_{12} < 2 \times 10^{-2}\;, &&\quad 
5\times
10^{-6} < \Delta m_{12}^2 < 10^{-5}\;. \label{eqconstraints}
\eea
The inverse neutrino mass matrix is 
\bea
m_{\rm eff}^{-1}&=&\pmatrix{1/m_1 & 0     & 0 \cr
                                      0 & 1/m_2 & 0 \cr
                                      0 & 0     & 1/m_3 }
\eea
This can be rotated by a mixing matrix
\bea 
V&=&V_{13}V_{12}V_{23}\;,
\eea
where $V_{ij}$ is a rotation matrix in between the $i,j$ rows and columns by
an angle $\theta_{ij}$. For example
\bea
V_{12}&=&\pmatrix{c_{12} & s_{12} & 0 \cr
                -s_{12} & c_{12} & 0 \cr
                      0 &    0   & 1 }\;,
\eea
where $s_{ij}=\sin \theta_{ij}$ and $c_{ij}=\cos \theta_{ij}$. We have 
taken the mixing matrix to be real for simplicity. Then the Majorana mass
matrix in this basis is
\bea
M_N&=&m_D^{\prime T} V m_{\rm eff}^{-1} V^T m_D^\prime \;,
\eea
The lepton asymmetry is created by the decay of the lightest of the
heavy majorana neutrinos. Consequantly we have to go to a basis in
which the Majorana mass matrix is diagonal. This can be diagonalized by a
matrix $K$ so that 
\bea 
M_N^{\rm diag}&=&K^T M_N K\;,
\eea
Note the Dirac and Majorana mass matrices are related through 
Eqn (\ref{eqn1}). It can be easily seen that the Dirac neutrino mass
matrix in such a basis is, 
\bea
m_D&=&K m_D^\prime  K^T\;.
\eea
The CP-asymmetries in the neutrino decays arize from the interference
between the tree level and one-loop level decay channels \cite{bp2,crv} 
\bea
\epsilon_j&=&-{1\over {8\pi v_2^2}}{1\over {(m_D^\dagger m_D)_{jj}}}
\sum_{n\ne j}{\rm Im} \left [(m_D^\dagger m_D)_{nj}^2\right ]
g\left ({{a_n}\over {a_j}}\right )\;,
\eea
where 
\bea
g(x)&=&\sqrt{x}\left [\ln \left ({{1+x}\over x}\right )+{2\over
{x-1}}\right]
\;,
\eea
and $v_2=v\sin \beta$.
The other parameter of most interest is the mass parameter 
\bea
\tilde{m}_1&=&{{(m_D^\dagger m_D)_{11}}\over {M_1}}\;, 
\eea
which largely controls the amount of dilution caused by the lepton number 
violating scattering \footnote{This parameter is especially important in
the supersymmetric scenarios there exists a large number of scattering
diagrams which are lepton number violating, and the Yukawa interactions are
much more important.}. 
A large enough lepton asymmetry can result only if $\tilde{m}_1$ is in 
the range $10^{-5} < \tilde{m}_1 < 10^{-2}$\cite{bp2}. For too
small values of $\tilde{m}_1$, the Yukawa interactions are too weak to
bring the neutrinos into equilibrium at high temperatures. For too high
$\tilde{m}_1$, the lepton number violating scatterings wash out most of
the asymmetry after it is generated.

Scanning over the alloowed ranges for the neutrino mixing angles  
and taking $\lambda=0.22$,
one can find the 
regions in the $m_1-m_2-m_3$ parameter space 
for which the two conditions are satisfied

\begin{itemize}

\item $|\epsilon_1| > 10^{-6}$

\item $10^{-5} < \tilde{m}_1 < 10^{-2}$

\end{itemize}

We assume for definiteness in our numerical results that 
the neutrino Dirac mass matrix is identical to the up quark mass 
matrix\footnote{Relaxing this choice, or choosing a different sign
for the neutrino mixing angles will change the quantitative results, but
not the qualitative ones.}.
There is an undertermined phase in this procedure which we can assume is 
such that the maximal CP-asymmetry is obtained since we are determining 
the points for which it is {\it possible} to obtain the required baryon 
asymmetry.

As an example the allowed regions for Texture 4 is shown in Fig.~1 with the
neutrino mixings set to $s_{23}=0.55$, $s_{12}=0.07$ and $s_{13}=-0.03$. 
For a particular choice of the neutrino mixing angles 
only a narrow three dimension region is 
allowed in the full parameter space. This region is characterized by 
larger values of the lightest right-handed neutrino mass 
($10^7 < M_1 <10^8$~GeV), and 
smaller values for the neutrino quark mass matrix parameter 
$(m_D^\dagger m_D)_{11}$. This represents a moderately fine-tuned solution
which can be understood from the hierarchical structure of the mass matrices
as follows.

\begin{center}
\epsfxsize=4.5in
\hspace*{0in}
\epsffile{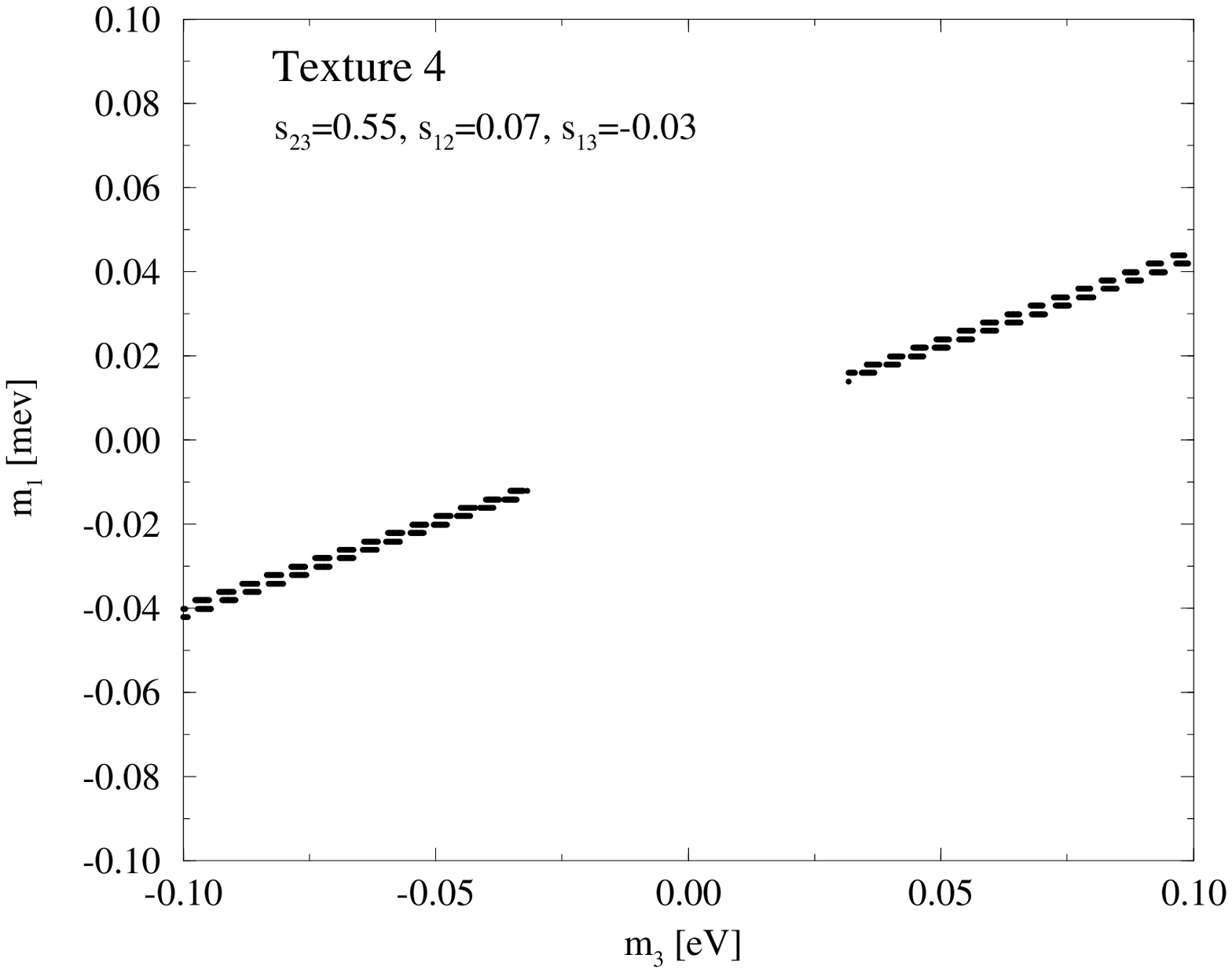}
\vspace*{0in}

\smallskip
\parbox{5.5in}{\small  Fig.~1. The allowed solutions for $m_1$ and $m_3$
that generate a sufficient baryon asymmetry for Texture 4. }
\end{center}
 
Consider a generic matrix exhibiting the hierarchy given by the textures 
\bea
\lambda _D^{\rm generic}&=&\pmatrix{A\lambda^8 & B\lambda^6 & C\lambda^4 \cr
                                    B'\lambda^6 & D\lambda^4 & E\lambda ^2 \cr
                                    C'\lambda^4 & E'\lambda^2   & 1 }\;,
\eea
where $A$, $B$, $B'$ \dots $E'$ are coefficients of order one.
The leading $\lambda$ dependence of the parameters most important to 
leptogenesis for this generic hierarchy is shown in Table~\ref{table2}.
Also shown are the $\lambda $ dependence of the five RRR textures; one sees 
that Textures 2, 4 and 5 have CP violation $\epsilon _1$ of the same order 
as the generic case, while Textures 1 and 3 are further suppressed by 
powers of $\lambda$ (the suppression by $\lambda ^4$ assumes that the 
texture zeros are exact). Furthermore Texture 1 has a enhanced $\tilde{m}_1$ 
which implies that for this texture the dilution factor is very large.

\begin{table}[htb]
\begin{center}
\[
\begin{array}{|c||c||c||c||c|}
\hline
Texture & (m_D^\dagger m_D)_{11} &  M_1 & \tilde{m}_1 & \epsilon_1 \\ 
\hline
&&&&\\
{\rm generic} & \lambda^{8} & \lambda^{16} & \lambda^{-8} & \lambda^{12} 
\\
&&&&\\
1 & \lambda^{12} & \lambda^{16} & \lambda^{-4} & \lambda^{16} 
\\
&&&&\\
2 & \lambda^{8} & \lambda^{16} & \lambda^{-8} & \lambda^{12} 
\\
&&&&\\
3 & \lambda^{8} & \lambda^{16} & \lambda^{-8} & \lambda^{16} 
\\
&&&&\\
4 & \lambda^{8} & \lambda^{16} & \lambda^{-8} & \lambda^{12} 
\\
&&&&\\
5 & \lambda^{8} & \lambda^{16} & \lambda^{-8} & \lambda^{12} 
\\
&&&&\\
\hline
\end{array}
\]
\end{center}
\caption{The leading $\lambda $ dependence of various parameters important 
to leptogenesis.
Textures 1 and 3 have leading $\epsilon_1$ that is subleading in $\lambda$,
so any solutions are exceedingly fine-tuned. In addition, Texture 1 typically 
has too large a dilution factor because $\tilde{m}_1$ is enhanced. 
Only Textures 2, 4 and 5 offer solutions that
are not overly fine-tuned.}
\label{table2}
\end{table}

If one assumes the hierarchy suggested by the RRR textures then there must be
some fine-tuning of the light neutrino masses to get a 
solution with a sufficient amount of leptogenesis. That is because the 
parameter $\tilde{m}_1$ which governs the dilution of the produced lepton 
asymmetry is too large for typical values of the 
light neutrino masses, related via
the seesaw as
\bea
&&\tilde{m}_1\sim \lambda ^{-8}{{v_2^2}\over {M_1}}\;.
\eea
The RRR textures fall into the category of the generic texture defined above, 
where the eigenvalues are in the ratio $1:\lambda^{-4}:\lambda^{-8}$. We find that
in this case there is some fine-tuning required to achieve the required 
amount of leptogenesis; in Fig. 1 there is only a small pencil-like region
which produces an adequate lepton asymmetry. For other choices of the neutrino
mixing angles there is a different linear correlation between the neutrino 
masses, but the same fine-tuning is required. This is because $\tilde{m}_1$ 
predicted by the RRR textures is typically too large by a factor $\lambda^4$ 
and the CP-violation $\epsilon _1$ is too small by the same factor. 
If instead a more modest hierarchy for the neutrino Dirac masses is 
assumed, say $1:\lambda^{-2}:\lambda^{-4}$, then one finds that 
a {\it typical} 
value of the light neutrino masses allowed by the solar and atmospheric 
neutrino experiments can generated the required level of leptogenesis.
This suggests that the leptogenesis occurs more naturally in cases with the
reduced hierarchy, and the RRR textures must be fine-tuned to achieve the
required leptogenesis.

Since leptogenesis can only occur after the end of inflation, the subsequent 
thermal production of massive gravitinos can occur. The gravitinos interact
weakly and the late decays of these can modify the observed abundances of 
light elements or overclose the universe\cite{grav1}. The right-handed 
neutrino mass $M_1$ in our numerical solutions 
is sufficiently light to admit a solution 
to the gravitino problem\cite{grav2}.

\section{Boltzmann Equations}

The size of the lepton asymmetry that results can be calculated using 
the full set of Boltzmann equations\cite{bp1}. These have been studied in
the scenarios where there are only the Standard Model particles\cite{luty}, 
but have become available recently in the full supersymmetric case as 
well\cite{bp2}. We consider the supersymmetric case here since the 
supersymmetric Yukawa interactions are sufficient to 
produce a thermal population of right-handed neutrinos after reheating (the 
nonsupersymmetric model requires the introduction of new 
interactions\cite{plu}). The Boltzmann equations become quite involved for the
supersymmetric case, where it is know that the dilution factor can be enhanced
over the non-supersymmetric case because of the enhanced effect of the 
Yukawa interactions. 

In principle what is desirable is to scan over all possible values for the 
masses and mixings of the neutrinos that are consistent with the solar and
atmospheric neutrino oscillation experiments, and to determine the viable
parameter choices. We do not do that here for three reasons: (1) computational
power is exhausted after a few points, as each solution of the Boltzmann 
equation for a parameter choice involves numerically integrating a set of 
differential equations each of which involves a further numerical integration
(this integration is needed to calculate the reaction density for the two
body scatterings which can occur over the full kinematic range), (2)
the exact equality between the neutrino and up quark Dirac matrices is 
probably only approximate, so our results must be considered qualitative only,
and (3) there is an unknown phase in the Dirac neutrino mass matrix that 
controls the amount of CP violation in the heavy neutrino decays, so one 
can only determine an upper bound on the amount of lepton asymmetry generated.
So we confine ourselves here to demonstrating that a particular parameter 
and texture choice can produce a baryon asymmetry consistent with observed
result
\bea
&&Y_B={{n_B}\over s}=(0.6 - 1)\times 10^{-10}\;,
\eea
where $n_B$ is the number density of baryons, and $s$ is the entropy density.
This quantity conveniently is insensitive to the dilution that comes about
from the expansion of the universe. Similar densities $Y_i$ can be defined
for all number densities $n_i$.

Figure 2 shows the evolution of the neutrino densities and the lepton asymmetry
as a function of the temperature $T$ through $z=M_1/T$ for Texture 4 with 
neutrino masses of $m_1=1.5\times 10^{-5}$~eV, $m_2=3.0\times 10^{-3}$~eV and 
$m_3=4.0\times 10^{-2}$~eV. This point is one of the 
allowed solutions shown in Fig.~1 that
satisfies the requirements for neutrino oscillations and for the requirements
on $\epsilon _1$ and $\tilde{m}_1$.
For these masses the right-handed Majorana mass is 
$M_1=2.9\times 10^7$~GeV. Assuming a maximal CP-violating phase, the
amount of CP-violation from the decays of the lightest Majorana neutrino 
is $\epsilon_1=-2.1 \times 10^{-6}$. These masses are 
consistent with the constraints from the solar and atmospheric neutrino 
experiments in Eq.~\ref{eqconstraints}. The evolution of the densities proceeds
to the right as the temperature of the universe decreases.
The figure shows the equilibrium density 
of the lightest Majarona neutrino $Y_{N_1}^{\rm eq}$ along with the computed
density $Y_{N_1}$. Nonzero asymmetries of lepton number from fermions
$Y_{L_f}$ and from scalars $Y_{L_s}$ develop, and change sign (hence the dip
in the figure), and finally asymptote to a constants for values of 
$z=M_1/T>3-4$. Sscattering processes involving exchange of supersymmetric 
particles enforce that 
$Y_{L_f} \approx Y_{L_s}$. Finally, for completeness, we show the  
scalar neutrino asymmetry for the supersymmetric partner to the lightest
Majorana neutrino $Y_{1-}=Y_{\tilde{N_1^c}}-Y_{\tilde{N_1^c}\dagger}$.
This asymmetry also changes sign before eventually vanishing for large values 
of $z$. The total density of scalar neutrinos 
$Y_{1+}=Y_{\tilde{N_1^c}}+Y_{\tilde{N_1^c}\dagger}$ is indistinguishable from 
$Y_{N_1}$ and is omitted from the figure.

\begin{center}
\epsfxsize=4.5in
\hspace*{0in}
\epsffile{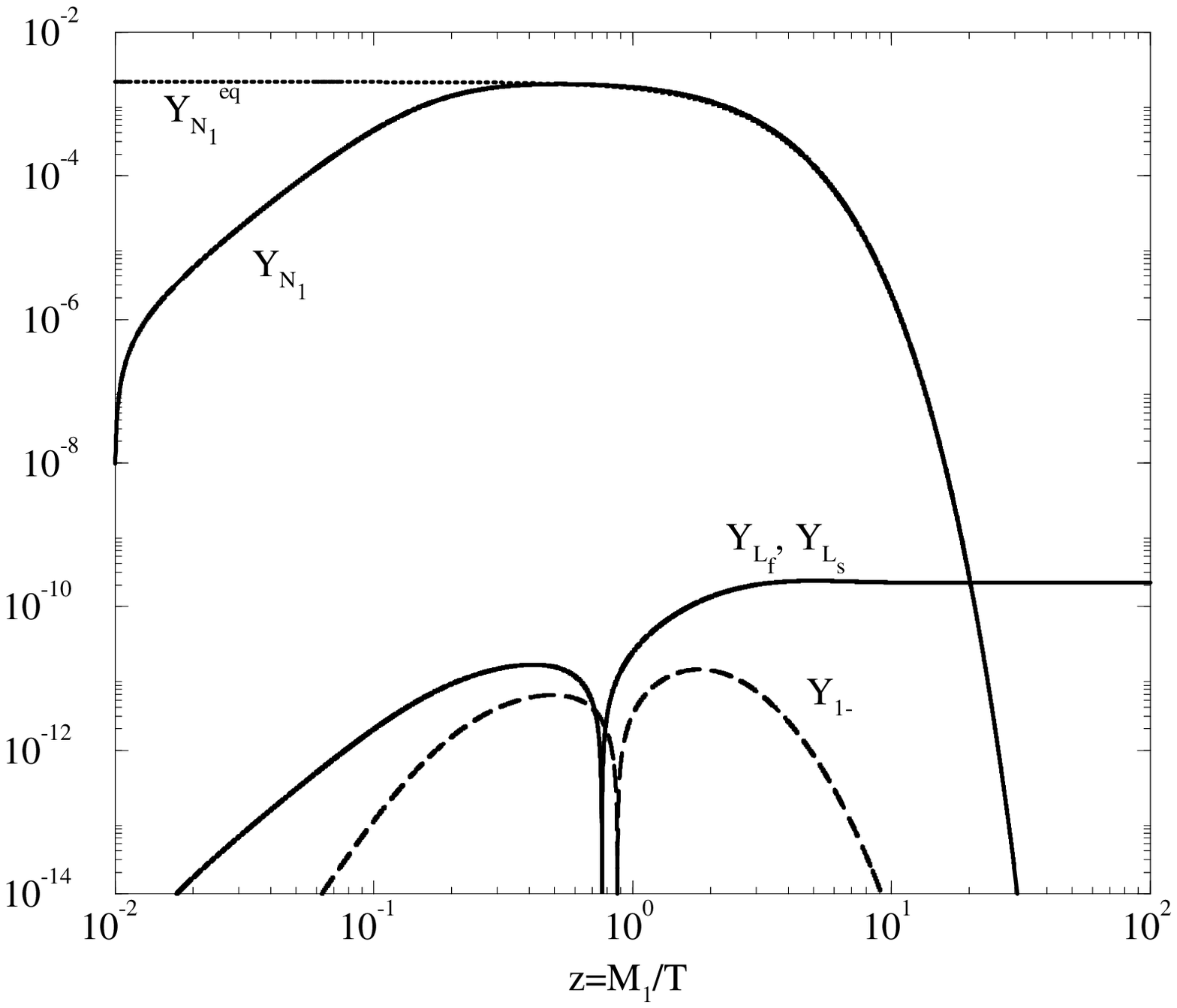}
\vspace*{0in}

\smallskip
\parbox{5.5in}{\small  Fig.~2. The evolution of the fermionic and baryonic 
lepton asymmetries. The asymmetries $Y_{L_f}$ and $Y_{L_s}$ 
asymptote to a constant value which is
recycled into a baryon asymmetry that is sufficient to account for 
experiment if 
$Y_{L_f}=Y_{L_s}=-(0.9-1.4)\times 10^{-10}$. }
\end{center}

The baryon asymmetry is related to the lepton asymmetry (in the supersymmetric
case) via
\bea 
&&Y_B=-{8\over 15}Y_L\;,
\eea
so the observed baryon asymmetry is generated provided that the asymptotic 
value (for small $T$) of the lepton asymmetry is 
\bea
&&Y_{L_f}=Y_{L_s}=-(0.6-0.9)\times 10^{-10}\;.
\eea
So Fig.~2 shows a consistent solution of Texture 4 that explains the baryon 
asymmetry of the universe which is arising after a fine-tuning of the neutrino 
masses.

\section{Conclusion}
We studied the possibility that the baryon asymmetry of the universe could
result from lepton number violating decays of heavy Majorana neutrinos. We
assumed the Dirac neutrino texture was given by the set of Ramond-Roberts-Ross 
textures with five zeros (which gives rise to correct masses and mixings
of the charged fermions). The heavy Majonana neutrino mass matrix is 
obtained by inverting the type-I see-saw formula where the contributions 
from the left handed triplet Higgs are neglected, and using the neutrino
masses and mixings required by the solar and atmospheric neutrino 
oscillation experiments. The lepton asymmetry is produced due to the
lepton number violating decay of the lightest right handed neutrino.
Contrary to naive expectations, the lightest eigenvalue of the heavy
Majorana neutrino mass matrix is in the range $10^5-10^7$ GeV
even though the right handed gauge symmetry breaks at $M_X=10^{16}$ GeV. 
This is due to the hierarchy 
of the Dirac-type neutrino texture. We obtained the
following results for the feasibility for each texture for generating the
required baryon asymmetry:
(a) A generic neutrino Dirac mass matrix with eigenvalues in the ratio
$1:\lambda^{-4}:\lambda^{-8}$ can produce the observed baryon asymmetry via
the baryogenesis via leptogenesis scenario in narrow ranges of 
light neutrino masses. This predicts a strong correlation between the light
neutrino masses, but only because the masses of the neutrinos must be carefully
tuned to achieve the required magnitude of leptogenesis.
A neuriino Dirac mass matrix eigenvalues in the ratio
$1:\lambda^{-2}:\lambda^{-4}$ 
naturally gives a lepton asymmetry of the required level.
(b) Textures 2,4,5 generate the amount of leptogenesis expected in models
with a neutrino Dirac mass hierarchy with eigenvalues in the ratio
$1:\lambda^{-4}:\lambda^{-8}$. (b) The position of the texture zeros 
in Textures 1 
and 3 result in a further suppression of the generated lepton asymmetry.

We carried out a detailed analysis of the generated baryon asymmetry by
solving the Boltzmann equations for a supersymmetric model numerically for
Texture 4. This demonstrates that the required texture can be compatible with
baryogenesis via leptogenesis for some specific values of the light 
neutrino masses consistent with observations in solar and atmospheric neutrino
experiments.

\end{document}